Learning Physics By Creating Problems: An Experiment

Ameya S. Kolarkar*, Aimee A. Callender**

*Department of Physics, Auburn University; **Department of Psychology, Auburn University

Abstract

We investigated the effects of student-generated problems on exams. The process was gradual with some training throughout the semester. Initial results were highly positive with the students involved performing significantly better, and showing statistically significant improvement ($t = 5.04$) compared to the rest of the class, on average. Overall, performance improved when students generated problems. Motivation was a limiting factor. There is significant potential for improving student learning of physics and other problem-based topics.

I.  **Motivation**

A common problem that ails non-major physics students (e.g. life science, pharmacy, architecture, building science) is the fear of the subject, whether it comes from their own perception from past experience or from others (Mallow, 1986; Udo, Ramsey, & Mallow, 2004). Compounding this issue is that a large percentage of our students have not taken physics in high school as data gathered through student surveys in the course indicate. Due to general fear of physics and a lack of experience with physics, students appear to enter into a stress/panic mode during exams causing them to perform even worse than they actually would have otherwise.

(They do better in active learning settings, until exam panic sets in.) The experiment described in this paper is an attempt towards alleviating the fear of Physics resulting in better assessment outcomes, and enhancing the understanding of content knowledge along the way.

Specifically, we investigated whether students making their own exam problems might improve exam performance. This method increases familiarity with problems and is not too complicated for this level. Also it gives them the opportunity to think deeply about the concepts which they typically would not when simply solving a given problem. By creating a problem, the student is engaging in a task at the top of the revised Bloom's Taxonomy pyramid (Anderson & Krathwohl, 2001).

## II.  Current Studies

This study was conducted in two separate classes of the first semester of the algebra-based introductory physics sequence for life science students. The first study was conducted in the fall semester and the second study was conducted in the spring semester.

To ensure that students were capable of creating questions, appropriate scaffolding was implemented.  Over the semester the students chose a few problems from their homework and make slight modifications ("mods") to those problems. Some mods were straightforward – backwards treatment of variables (inverting given and wanted)- while others were indeed highly thought out, elaborate problems which were difficult even by the instructor's expectations for the class.  Some simply replaced surface features, e.g. a "bike" from the homework problem to "car" but this was nevertheless deemed significant given the students' prior perceptions about "never-before-seen" exam problems.

## III. Study 1: Fall 2014

### a. Methods

*Participants*. The course was a 1st year physics course for non-majors. There were 73 students in the course that participated in the study, and they came from a variety of majors, including: pre-med and pre-vet (~65%), pre-pharm (~31%), architecture and building science (<5%).

*Design and Procedure*. There were a total of four exams this semester (50 minute class, 3 days a week). The first exam was a regular exam with the instructor choosing all of the questions for the exam. It had problems taken from the homework but certain surface features of the problem were altered. The second exam was just like the first one – it was used to determine if the first exam performance was reliable.

Exam 3 is when this experiment began in a scaffolded form. For exam 3 the students were asked to vote on homework problems that they would like to see on the exam. For exam 3, there were three problem sets to vote on. Students were also allowed to choose how many problems they wished to solve for their 80% of the exam. 66% voted for 3 problems, 33% voted for 4 problems, 1 student voted for 2 problems[4]. So the exam structure was 3 of their problems for 80% of the exam score, and instructor's choice for the remaining 20%. About 90% of the students consistently voted on the problems.

Each homework set had 12 problems that covered a chapter from Serway-Vuille, College Physics 10e. The platform used to vote for problems was Piazza, where the students could vote anonymously. In each homework set vote, there was a clear set of 5 problems with a statistically significantly higher number of votes than the other problems in that homework set. The five problems with the highest number of votes were included in a test bank. The students knew what these problems were. Exam questions were taken directly from the test bank with no changes whatsoever for the portion of the test that students voted on which accounted for 80% of the exam grade. The remaining questions worth 20% were chosen by the instructor, and were complementary to the ones chosen by the students so as to include all topics covered on that test. (As an interesting aside, the students chose the easy problems from the easy topics leaving the difficult topics to the instructor. They quickly realized their folly after the test and made proper choices thereafter.)

---

[4] On interviewing the one student who chose 2 problems, she revealed that she thought it would give her more time to work out the other 20% of the test which she assumed would be difficult. (She scored an A in the course.)

For exam 4, students were given the opportunity, but were not required, to modify the 5 most-voted-for exam problems. (This is the subject of discussion in this paper.) For each of exams 3 and 4, students were given a minuscule points incentive to vote on the problems.

For exam 4, the logistics were the same as those for exam 3 except for the content, and the fact that they were asked to make modifications to the problems. About 90% students voted on the problems but only about 10% made modifications or created their own problems.

The students who made modifications ("modders") were asked to post their mods or creations on Piazza so the other students could have access to the modifications. The modders handed in all their modifications to the instructor on paper but some of them put only a subset of their problems on Piazza. In total, about 10% of the class modified problems. These students consisted of students who had failed the first exam as well as a couple who received A's on the first exam.

b. **Results**

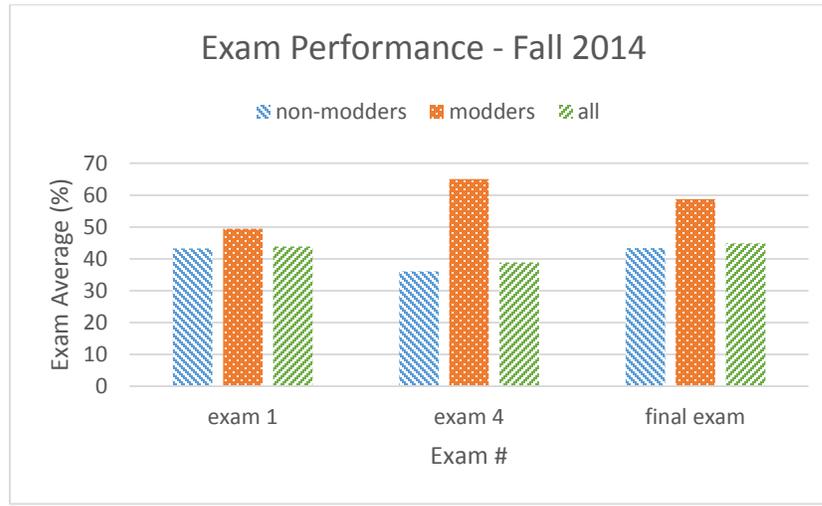

*Figure 1: Student performance on tests 1, 4 and Final. Tests 2 and 3 were part of the scaffolding process and not included in the analysis.*

Two two-tailed t-tests were conducted to compare performance between the students who completed modifications and those who did not.  The first t-test compared performance on Exam 1 (before the modifications were implemented) and the second t-test compared performance on Exam 2 (after modification were made).  For exam 1, there was no difference in test performance between those who later made modifications and those who did not, $t(70) = 0.68$, $p = 0.55$. Therefore, prior to the modification task was implemented, there were no differences in test performance between the groups.  On Exam 4, the modders performed much better on the student-sourced exam (exam 4) than those who did not, as is evident in the graph in Figure.  A t-test was conducted to compare test performance between those who completed the modifications and those who did not on exam 4 performance, $t(70) = 5.04$, $p = 0.00088$, and the modders performed significantly better than the non modders.

Further, critical understanding was tested using problems assessing concepts of Angular Momentum. Angular Momentum final exam average was about 58% for the rest of the class versus 90% for the modders. Informal interviews revealed impressive levels of understanding among the modders.

## IV. Study 2: Spring 2015

### a. Methods

Inspired by the positive result from the Fall 2014 semester, we decided to conduct a much more controlled experiment the following Spring. The goal was to assess quantitative gains in student understanding using this method. But this semester was much more interesting and unexpected.

*Participants.* Students in the same course – algebra-based physics 1 – in a similar classroom setting were given the same options as the ones in Fall 2014. These were about 160 students again mostly from the life sciences, architecture and building science.

*Design & Procedure.* During this semester, the first exam was given as a baseline, with no voting or modifications. Then, we attempted to tease apart the effects of voting vs. making modifications. For exam 2, students were encouraged to vote for the problems, and extra credit was given for voting. For exam 3, modifications were encouraged, and again, extra credit was provided for making modifications.

### b. Spring 2015 Results

To our dismay, even with the extra credit that was provided, the students were not motivated enough to participate in the activity. A lot of encouragement and pleading by some other students was necessary for the number of students voting to cross 50% for both exams 2 and 3. Similar was the state of modifications for exam 3 where a handful (about 5% out of a total of about 160 students) frantically submitted their "modifications" a day or two before the exam. These were not at all thought out and borne out of desperate attempt to simply get the extra credit. The exam averages were in the high 50s this time but low enough to scare the students.

## V.   Discussion

The inconsistency of the results between the fall and spring semesters make it difficult to draw strong conclusions about the method.  When students take advantage of the method, it seems to be very effective.  However, motivating students to modify questions proved to be difficult.  It was surprising to us that the students did not take up the opportunity to succeed through this avenue. Their non-motivation to succeed or perform better is still a mystery to us.

We have considered several psychological factors that could have contributed to the students' choices.  First is the concept of delay discounting, or that people discount the value of a reward based on the temporal delay to receive the reward.  If students do not receive an immediate reward for completing a task, they may not do it.  In this scenario, in the fall semester the benefits of completing the modifications were not received until the exam which could have been several weeks later.  We attempted to institute a more immediate reward (extra credit for submitting modifications in the spring semester), but that was not enough of an incentive for the students to create the modifications.  Therefore, we believe that two other factors were at play that substantially decreased motivation:  self-efficacy and spatial processing ability.

Self-efficacy is the belief that an individual holds about their ability to complete a task and is related to their locus of control, or whether they believe they have control over the outcomes of events (Bandura, 1977), (Judge, Erez, Bono, & Thoresen, 2003). Students with low self-efficacy may believe that there is nothing they can do to improve performance in physics class, reducing the probability that they would complete the modifications. Low self-efficacy may be a significant factor for students with low spatial processing ability. The ability to visualize is important to solving problems in physics. Students who are categorized as having low spatial ability fail to integrate or combine vectors, interpret kinematics graphs as pictorial representations of movement, and fail to integrate different components of a problem into a single, cohesive, representation (Kozhevnikov, Motes, & Hegarty, 2007). Students who have difficulty with visualization and have low self-efficacy may have a high degree of science, specifically physics, anxiety. These barriers may prevent them from taking advantage of methods that could improve their performance.

## VI. Transfer

Transfer is a big problem area in Physics, and in the sciences in general. We anticipated that modifying existing questions would lead to greater transfer and better learning overall. They have already worked so hard coming up with their problems that they have already learned more than their peers, and more than they would have themselves otherwise. They were naturally also presumed to do better on the exams than their peers to the point of rendering the standard assessments moot. These students were assumed to have achieved, as determined anecdotally/qualitatively, the *learning objectives* for the course.

And indeed this is what was observed. The modders were informally interviewed and the clear conclusion was that they really do understand the concept in depth. And when confronted with a different and difficult scenario, they were able to come up with the solution all by themselves after some thought. Also, they did not give up, and were able to correlate their knowledge from other scenarios that they already knew or had thought about (outside of what was done in class) to solve the new problem. It was impressive. There was *transfer*.

It is possible that the positive effects that were observed were simply due to increased time-on-task. It could be argued that through the "creating" process, the students had more time on task which has been shown to improve outcomes. Nevertheless, this could be considered a positive and a motivation for students to spend more time on task. However, the depth of understanding as observed through individual interviews reveal a greater improvement than pure time on task. This will need to be systematically tested.

**VII.    Future Directions**

In the Fall 2015 and Spring 2016 semesters the modification and creation process has been moved to the recitation sections for both of the instructor's algebra-based Physics 1 classes. This relies heavily on the Graduate Teaching Assistants, but Undergraduate Learning Assistants have also been implemented. Both TAs and LAs are being trained and discuss various recitation strategies weekly. This implementation has seen anecdotal success in that the students are using their imagination to come up with brilliant modifications to existing problems, as well as creating their own. They are able to solve their own mods as well as the ones by other groups, and they are able to identify when information is missing from some problems. This is a dramatic improvement in the attitude of these students towards physics as a whole, and the entire dynamics of the recitation sessions is positive and fun.

**VIII.    Summary**

Requiring students to make modifications to existing problems and creating new problems can have dramatic effects on performance.  The difficulty, however, is in the implementation of the strategy due to low levels of motivation. However, incorporating the method in recitation sections that are supported by GTAs and LAs has shown initial success in terms of participation and motivation. Future research will have to investigate outcomes from classes in which this implementation is used.